\documentclass[aps,psfig,showpacs]{revtex4}

\usepackage{epsfig}

\usepackage{amsmath}

\begin{document}
\draft
\title{Asymptotically exact trial wave functions for yrast states of rotating Bose gases}
\author{S. Viefers and M. Taillefumier}
\address{Department of Physics,  University of Oslo,
P.O. Box 1048 Blindern, N-0316 Oslo, Norway }
\date{\today}
\begin{abstract}

\noindent
We revisit the composite fermion (CF) construction of the lowest angular momentum yrast states
of rotating Bose gases with weak short range interaction. For angular momenta at and below the single vortex, $L \leq N$, the 
overlaps between these trial wave functions and the corresponding exact solutions {\it increase} with increasing system size and 
appear to approach unity in the thermodynamic limit. In the special case $L=N$, this remarkable behaviour was previously observed numerically.
Here we present methods to address this point analytically, and find strongly suggestive evidence in favour of similar behaviour for all  $L \leq N$. While not constituting a fully conclusive proof of the converging overlaps, our results do
demonstrate a striking similarity between the analytic structure of the exact ground state
wave functions at $L \leq N$, and that of their CF counterparts. Results are given for two different projection methods 
commonly used in the CF approach.

\end{abstract}
\maketitle
\vspace{-11pt}

\newcommand{\half}{\frac 1 2 }
\newcommand{\eg}{{\em e.g.} }
\newcommand{\ie}{{\em i.e.} }
\newcommand{\etc} {{\em etc.}}

\newcommand{\noi}{\noindent}
\newcommand{\etal}{{\em et al.\ }}
\newcommand{\cf}{{\em cf. }}

\newcommand{\dd}[2]{{\rmd{#1}\over\rmd{#2}}}
\newcommand{\pdd}[2]{{\partial{#1}\over\partial{#2}}}
\newcommand{\pa}[1]{\partial_{#1}}
\newcommand{\pref}[1]{(\ref{#1})}

\newcommand{\be}{\begin{eqnarray}} 
\newcommand{\ee}{\end{eqnarray}}
\newcommand{\e}{\varepsilon} 
\newcommand{\D}{\partial}
\newcommand{\pt}{\tilde p}
\newcommand{\zt}{\tilde z}
\newcommand{\zb}{\bar z}
\newcommand{\p}{\partial}
\newcommand{\jas}{\prod_{i<j}(z_i - z_j)}

\section{Introduction}

\noi
The behaviour of ultracold atomic Bose gases under rotation has been a subject of intense study over the past years\cite{fetter}. One of the most studied aspects, involving a large body of theoretical work, is the expected occurrence of strongly correlated states of the fractional quantum Hall (FQH) type in the limit of ultrafast rotation\cite{reviews}. On the experimental side, progress has been extremely impressive, from the creation of the first vortex in 1999\cite{svor} to arrays of hundreds of vortices\cite{mvor}. Nevertheless there are practical obstacles to reaching the actual quantum Hall regime this way. Other, more promising, scenarios are now being pursued to try and create the effective 'magnetic' fields required for producing bosonic FQH states\cite{synthetic}.
Another limit that has attracted considerable interest is that of {\it slow} rotation, in the sense that only one or a few vortices are present in the system. In particular, various groups have studied the {\it yrast line}\cite{mottelson} of the system, i.e. the states with lowest energy for a given total angular momentum. Theoretical work includes analytical studies\cite{mottelson}, exact diagonalization\cite{matti}, the Gross-Pitaevskii approach\cite{kavoulakis2000, lieb}, as well as trial many-body wave functions\cite{cooper1,viefers1}. Moreover, for the lowest angular momentum regime, $2 \leq L \leq N$, exact analytical ground state wave functions were found for bosons with short-range interaction sufficiently weak to assume the particles reside in the lowest Landau level\cite{wilkin1,bertsch99,smith1,jackson,papenbrock01}. 

The focus of the present paper is this latter case. The composite fermion (CF) approach\cite{jainbook}, originally designed to give trial wave functions in the fractional quantum Hall regime, was first applied to some low-angular momentum states in Refs.\cite{cooper1,viefers1}. It was noted\cite{viefers1} that overlaps between CF and exact wave functions for the 'single vortex' state at $L=N$ appeared to increase with particle number, at least for small systems. This point was studied more systematically in Ref.\cite{korslund}, and overlaps for the single vortex state computed numerically for up to 45 particles. The results strongly suggested convergence of the overlaps to unity in the thermodynamic limit, with deviation decreasing as $\sim 1/N$. This kind of behaviour is generically unusual for {\it any} kind of non-exact trial wave functions. In this case it was even more surprising because {\it a priori}, the composite fermion construction could not be expected to work well in this low angular momentum regime -- for reasons to be explained below. Unfortunately, the numerical calculations gave little insight into the reasons for this behaviour. Therefore, we here reexamine the issue analytically, including also the states below the single vortex: We study the yrast states in the entire regime $2 \leq L \leq N$, comparing the analytic form of the CF wave functions to their exact counterparts, and present evidence that the increase of overlaps with system size occurs for all these states. We address the issue for two different projection methods commonly used in the CF approach. In the case of 'full projection' (method I) we demonstrate that the analytic structure of the CF wave functions is strikingly similar to that of the exact ones; the latter have the form of fundamental symmetric polynomials in the coordinates $(z_i - Z)$, where $Z$ is the center of mass, while the former are merely 'lacking' one or several coordinates in the center of mass term. For the other projection method (method II), it is possible to rewrite the CF wave functions, as well as the exact ones, in the form of an $N \times N$ determinant, divided by a Jastrow factor. We analytically compute overlaps between pairs of corresponding {\it entries} in these determinants and show that {\it the overlaps between all pairs of entries converge to unity in the thermodynamic limit}. This does not necessarily allow us to conclude that the overlaps between the full determinants (and thus the wave functions) go to unity as $N \rightarrow \infty$, since the size of the matrices themselves grows with $N$; the latter remains a challenging open problem.

Although exact analytic yrast wave functions are known for the particular states addressed in this paper, it is nevertheless of interest to examine why the CF trial wave functions work so surprisingly well in this regime.
In particular, this may help justify applying the CF construction to nearby states where no exact solutions are known -- such as excitations above the present yrast states, or other few-vortex yrast states at $L > N$.

The outline of the paper is as follows: Section \ref{sec:bg} presents some necessary background on rotating bosons in the lowest Landau level, the form of the exact yrast wave functions for $2 \leq L \leq N$, as well as the composite fermion approach. Sections \ref{sec:svp1} and \ref{sec:svp2} contain results for the single vortex state $L=N$ for the two different projection methods. The methods developed here can be directly transferred to all the lower angular momentum states, $L < N$; this is the topic of section \ref{sec:LlN}. Section \ref{sec:concl} gives some concluding remarks. An appendix is included, providing some details of the overlap calculations in section \ref{sec:svp2}.

\section{Background}
\label{sec:bg}

\subsection{Rotating bosons in the lowest Landau level}

\noi
Consider a system of $N$ spinless bosons with mass $m$
in a harmonic trap of strength $\omega$,
rotating with angular frequency $\Omega$ 
and interacting via a short-range (delta function) potential $H_I$. In a rotating frame
the Hamiltonian can be written as
\be
H = \sum_{i=1}^N \left[\frac{\vec p_i^2}{2m} + \frac 1 2 m \omega^2 \vec r_i^2 \right]
-\Omega L_z + H_I
\ee 
where  $L_z$ denotes
the angular momentum around the rotation axis.
Completing the square inside the brackets, this may be recast as 
\be
H = \sum_{i=1}^N \left[ \frac{1}{2m}\left( \vec p_i - \vec A \right)_{\parallel}^2  
+ H_{ho}(z_i)\right] 
+(\omega -\Omega) L_z + H_I
\ee
with $\vec A = m\omega (-y,x)$, $H_{ho}(z)$ denoting the $z$-part of the harmonic 
oscillator Hamiltonian and $\parallel$
denoting the planar ($x,y$) part of the Hamiltonian. We note that the planar part of $H$
takes the form of particles moving in an effective  "magnetic" field
$\vec B_{eff}=  \nabla \times \vec A = 2m\omega \hat z$.

Now, the interaction is assumed to be weak in the sense that it does not mix different
harmonic oscillator levels. We will be interested, for a given total angular momentum, 
only in the {\em lowest} many-body states (the "yrast" states).
In this limit, the model may be rewritten as a lowest
Landau level (LLL) problem in the effective "magnetic" field
$B_{eff}= 2m\omega$ (and of course, $n_z=0$ for the harmonic oscillator in the
 $z$-direction). The Hamiltonian then takes the form
\be
H = (\omega - \Omega)  L + g\sum_{i<j} \delta^2(\bf r_i - \bf r_j)
\ee
($\hbar=1$),
where we now use $L$ to denote the total angular momentum,
$L=\sum_i l_i = L_z$.

In the following we shall be concerned with many-body wave functions describing this system
at some given number of particles and total angular momentum.
A convenient basis of single particle states spanning our Hilbert space (the lowest Landau level)
is given by
\be
\eta_{0,l} = \frac{1}{\sqrt{2^{l+1}\pi l!}} z^l e^{-\bar z z/4}
\ee
where $z = \sqrt{2m\omega} (x+iy)$ are complex coordinates denoting the
particle positions in the plane, and $l$ is the angular momentum of the state.
A general bosonic many-body wave function $\psi(z_1,...z_N)$ 
will then be a homogeneous, symmetric polynomial in the $z_i$'s, times the
exponential factor $\exp(-\sum_i |z_i|^2/4)$ 
(which will be suppressed throughout most of this paper for simplicity). The degree of the
polynomial gives the total angular momentum of the state.

In connection with the CF construction, we shall need more general single particle basis states for
the full Landau problem, {\it i.e.} all Landau levels. In symmetric gauge, these are given by
\be
\eta_{n,m} = N_{n,m} \, e^{-|z|^2/4} \, z^m L_n^m \left( \frac{z\bar z }{2} \right),
\label{llwf}
\ee
where $n$ is the Landau level index, $m$ denotes the angular momentum,
$N_{n,m}$ is a normalization factor, and $L_n^m$ are the associated Laguerre
polynomials.

\subsection{Exact ground states}

\noi
Here and in the following we will use the notation $\psi_{[L,N]}$ for the (ground state) wave function of an $N$-particle state at total angular momentum $L$. Exact LLL ground state wave functions for the Hamiltonian discussed above are known for all angular momenta $2 \leq L \leq N$. As was
shown in a series of papers\cite{wilkin1,bertsch99,smith1,jackson,papenbrock01}, the yrast states in this angular momentum interval are simply 
given by fundamental symmetric polynomials ${\cal S}_L(\tilde z_i)$ where 
$\tilde z_i = z_i - Z$, and $Z=(\sum_i z_i)/N$ is the center-of-mass coordinate:
\be
\psi_{[L,N]}^{ex} = \sum_{p_1 < p_2 < ... < p_L} (z_{p_1} - Z)(z_{p_2} - Z) \cdots (z_{p_L} - Z).
\label{TI1}
\ee 
For example, $\psi_{[2,N]} = {\cal S}\left[(z_{1} - Z)(z_{2} - Z)\right] $, with $\cal S$ denoting
symmetrization over all particle coordinates. As a special case the so-called single vortex state
at $L=N$ obeys the exact ground state wave function
\be
\psi_{[N,N]}^{ex} = \prod_{i=1}^N (z_i - Z).
\label{svex}
\ee

\subsection{Composite fermion approach}

\noi
The phenomenology of {\it composite fermions}\cite{jainbook} was first introduced in the context of the fractional quantum Hall effect (FQHE), where it was used to construct very successful trial lowest Landau level many-body wave functions for a large number of FQH states. Roughly speaking, the basic idea of this construction is that an even number of vortices is bound to each electron. Each such vortex, in a mean field sense, cancels one flux quantum of the external magnetic field, so that the strongly interacting electrons are mapped to weakly interacting {\it composite} fermions, moving in a reduced effective magnetic field. A mathematical way of seeing that these composite objects are weakly interacting is that 'attaching $p$ vortices' amounts to multiplying the many-body wave function by $\prod_{i<j} (z_i - z_j)^p$ -- a factor that 'keeps the particles apart', or introduces an additional $p$-fold correlation hole around each particle, since it can be seen to go to zero as any two particle coordinates approach each other. A modified version of this model was later applied to study quantum Hall-type states in rapidly rotating Bose gases, mapping the lowest Landau level bosons to composite fermions by attaching an {\it odd} number of vortices\cite{cooper1,viefers1,chang1,regnault1,reviews}. Due to the above qualitative picture of composite fermions being weakly interacting objects, they are commonly approximated as non-interacting. Trial wave functions for LLL bosons at some angular momentum $L$ are thus constructed as a Slater determinant $\Phi$ of composite fermions with angular momentum $L-pN(N-1)/2$, times an odd power $p$ of Jastrow factors,
\be
\psi_{[L,N]}^{CF}(\{ z_i \}) = {\cal P}\, \left( \Phi(\{ z_i, \bar z_i \}) \, \prod_{i<j}^N (z_i - z_j)^p \right).
\label{cfwf}
\ee
The operator ${\cal P}$ projects the wave function to the lowest Landau level. This is necessary since, in the intermediate step, the Slater determinant $\Phi$ typically contains particles occupying states in higher Landau levels in the {\it effective} magnetic field, cfr. Eq.\pref{llwf}. Essentially, the projection amounts to replacing $\bar z_i$ by $\p / \p z_i$ in the polynomial part of the wave function. In the quantum Hall literature there are several ways of doing this in practice, and numerically the results are known to depend very little on the projection method chosen\cite{jainbook}. We shall get back to the details of this in later sections. However, let us point to one important issue already now: Assume the Slater determinant $\Phi$ is such that a majority of the composite fermions reside in higher effective Landau levels. This would translate to large powers in $\bar z$ (see Eq.\pref{llwf}) and thus, upon projection, to a large number of derivatives acting on the factor $ \prod_{i<j} (z_i - z_j)^p$. Given the above qualitative arguments, if the number of derivatives is sufficiently large compared to the degree of the polynomial itself, one might expect that these derivatives would destroy the good correlations built in by the 'vortex attachment', so that the approximation of non-interacting composite fermions underlying \pref{cfwf} might be rather poor. 
In particular, naively applying the recipe \pref{cfwf} to the lowest angular momentum states at $L \leq N$ involves ${\cal O}(N^2)$ derivatives acting on an order $N^2$ polynomial in $z_i$, killing off basically the entire vortex attachment factor, to leave a polynomial of degree $N$ or less. 
So these states are far outside the quantum Hall regime, and one would not, {\it a priori}, expect the CF scheme to work very well. Nevertheless one finds the very surprising fact that the overlap of the composite fermion trial wave functions with the exact ones \pref{TI1} {\it increases} with increasing particle number (being very large already for small systems), and approaches 1 as $N \rightarrow \infty$ -- in other words, the composite fermion wave functions appear to be analytically exact in the thermodynamic limit. For $L=N$, this trend has previously been pointed out in the context of numerical calculations\cite{viefers1,korslund}. In the present paper we study the analytic structure of the yrast wave functions in the entire interval $2 \leq L \leq N$, and present analytic evidence of how this highly unusual behaviour comes about.

\section{Single vortex -- projection I}
\label{sec:svp1}

\begin{figure}
{\psfig{figure=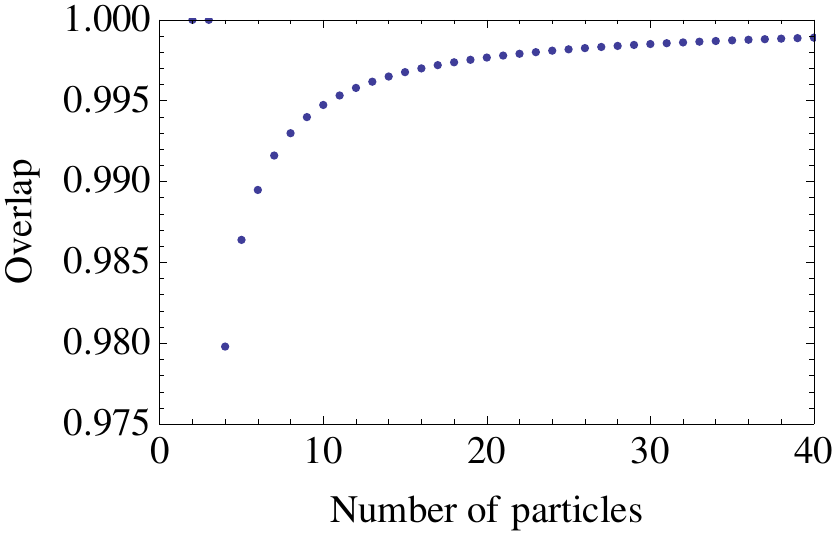, scale=1.0,angle=-0}}
\caption{Numerically computed overlap between exact and CF single vortex wave function for projection method I\cite{korslund}.}
\label{fig:nico}
\end{figure}

\noi
With the preliminaries in place, we are ready to study the single vortex state, $L=N$. For 'full projection' (referred to here as projection method I and explained below), this case was previously studied numerically\cite{korslund}; in Fig.\ref{fig:nico} we see numerically computed overlaps between the composite fermion and exact single vortex wave functions for up to 40 particles. This figure shows how the overlap converges towards unity, which is quite unusual behaviour for trial wave functions in general. As argued above, the fact that we are far outside the regime where the CF ansatz is naively expected to work, makes this result even more surprising. 
A full analytic proof for the convergence of these overlaps is still missing. But in this section we show how the CF wave function may be recast in a closed form, demonstrating that the analytic structure of the Jain states is extremely similar to that of the exact ground state; the main difference lies in that the center of mass coordinate in the exact wave function is replaced by 'incomplete' center of mass factors, with one or more particle coordinates 'missing'.
The results of the present section can be directly used to find similar, closed expressions for the lower angular momentum Jain states, $L<N$. These results will be presented in section \ref{sec:LlN}. 

\noi
According to Eq.\pref{svex}, the exact ground state wave function at $L=N$ is given by 
$\prod_i(z_i - Z)$ where the center of mass coordinate $Z = \sum_i z_i /N$.
The bosonic CF ansatz for the single vortex state, on the other hand, is given by\cite{viefers1,korslund}
\be
\psi_{[N,N]}^{CF}(\{ z_i \}) = {\cal P}\, \left( \Phi_{[N,N]}(\{ z_i, \bar z_i \}) \, \prod_{i<j} (z_i - z_j) \right),
\ee
where the unprojected CF Slater determinant is
\be
\Phi_{[N,N]}(\{ z_i, \bar z_i \}) =  \left| \begin{array}{cccc}
z_1 & z_2 & ... & z_N \\
1 & 1 &  ... & 1 \\
\zb_1 & \zb_2 & ... & \zb_N \\
\zb_1^2 & \zb_2^2 & ... & \zb_N^2 \\
... & ... & ... & ... \\
... & ... & ... & ... \\
\zb_1^{N-2}& \zb_2^{N-2} & ... & \zb_N^{N-2}
 \end{array} \right|.
\ee
Now, what we call projection method I, or full projection, simply amounts to replacing all $\bar z$'s in the Slater determinant by derivatives, $\zb_i^k \rightarrow \p_i^k$, thus acting on the full Jastrow factor to its right. We now present an iterative method by which we can perform all the derivatives to find closed, analytic expressions for any number of particles. To this end, first note that for $L=N=3$ we know from translation invariance arguments\cite{korslund} that the CF ansatz has to be identical to the exact wave function,
\be
\left| \begin{array}{ccc}
\p_1 & \p_2 & \p_3  \\
1 & 1 &  1 \\
z_1 & z_2 & z_3
 \end{array} \right| \, 
 \prod_{i<j}^3 (z_i - z_j )= \left( z_1 - Z \right) \,  \left( z_2 - Z \right) \, \left( z_3 - Z \right) \, ,
 \label{LN3}
\ee
with $Z = (z_1 + z_2 + z_3)/3$, as can of course be checked explicitly. This will help us simplify the corresponding expressions for larger $N$. To see how this works, consider the case $L=N=4$ and expand the $4\times 4$ Slater determinant by the row with the largest number of derivatives,
\be
\psi_{[4,4]}^{CF} = \sum_{i=1}^4 (-1)^{i+1} \left| \begin{array}{ccc}
\p_k & \p_l & \p_m  \\
1 & 1 &  1 \\
z_k & z_l & z_m
 \end{array} \right|^{(i)}  \p_i^2 \prod_{\alpha < \beta}^4 (z_{\alpha} - z_{\beta}),
\label{LN4a}
\ee
where the superscript $(i)$ on the Slater determinant denotes that coordinate $i$ is excluded. Next, note that the Jastrow factor in \pref{LN4a} may be split up according to
\be
\prod_{\alpha < \beta}^4 (z_{\alpha} - z_{\beta}) = \prod_{\alpha < \beta}^{~~~~~(i)} (z_{\alpha} - z_{\beta}) \cdot (-1)^{i+1} \prod_{k \neq i}^3 \left( z_i - z_k \right),
\label{LN4b}
\ee
where the sign factor comes from moving coordinate $i$ to the front. Furthermore, up to irrelevant overall multiplicative factors,
\be
\p_i^2 \prod_{k \neq i}^3 \left( z_i - z_k \right) \propto \left( z_i - Z \right) \propto  \left( z_i - Z^{(i)} \right), 
\label{LN4c}
\ee
where $Z$ is the center of mass of all four particles, while $Z^{(i)} = \sum_{j \neq i} z_j/3$ is the three-particle center of mass with coordinate $i$ omitted. Combining Eqs. \pref{LN4a}, \pref{LN4b} and \pref{LN4c}, we find
\be
\psi_{[4,4]}^{CF} = \sum_{i=1}^4  \left| \begin{array}{ccc}
\p_k & \p_l & \p_m  \\
1 & 1 &  1 \\
z_k & z_l & z_m
 \end{array} \right|^{(i)}  \prod_{\alpha < \beta}^{~~~~(i)} (z_{\alpha} - z_{\beta}) \cdot \left( z_i - Z \right) .
 \label{LN4d}
\ee
The derivatives in the reduced Slater determinant might in principle act through the Jastrow factor, onto the factor $\left( z_i - Z \right)$. However, it can be shown that any such contribution vanishes due to antisymmetrization. Thus, we identify the first part of \pref{LN4d} as the three-particle single vortex wave function \pref{LN3}, giving
\be
\psi_{[4,4]}^{CF} = \sum_{i=1}^4   \left( z_i - Z \right) \cdot \psi_{[3,3]}^{(i)}
\label{LNe}
\ee
or, equivalently (again neglecting irrelevant overall multiplicative factors),
\be
\psi_{[4,4]}^{CF} = \sum_{i=1}^4   \prod_{k=1}^4 \, \left( z_k - Z^{(i)} \right).
\label{LNf}
\ee
Note that, except for the one missing coordinate in the center of mass, this is identical to the exact ground state wave function \pref{svex}. If this was the case at all $N$, it would thus not be very surprising that the CF wave function approached the exact one in the thermodynamic limit. However, as we shall see now, going to higher particle numbers introduces factors where more and more coordinates are missing from the center of mass, as compared to the exact state. Thus, the convergence of the CF state to the exact one appears to be quite subtle.\footnote{Another way that this convergence might occur trivially, would be that the CF and exact wave functions simply shared the same leading term, and all other terms became irrelevant in the large $N$ limit. That this is not the case, was discussed in Ref.\cite{korslund}.}

The calculation outlined above for four particles, can be immediately iterated to larger systems. For example, for six particles,
\be
\psi_{[6,6]}^{CF} &=& \sum_{i=1}^6   \left( z_i - Z \right) \cdot \psi_{[5,5]}^{(i)CF} \\
                              &=&  \sum_{i=1}^6  \left( z_i - Z \right) \, \sum_{j \neq i}^5  \left( z_j - Z^{(i)} \right) \, \sum_{k \neq i,j}^4  \left( z_k - Z^{(ij)} \right) \, 
                              \prod_{l \neq i,j,k}   \left( z_l - Z^{(ijk)} \right).
\label{LN6}
\ee
It is straightforward to derive the general result
\be
\psi_{[N,N]}^{CF} &=& \sum_{i=1}^N   \left( z_i - Z \right) \cdot \psi_{[N-1,N-1]}^{(i)CF} \, ,
\label{LNgen}
\ee
which can be expanded iteratively in analogy with Eq.\pref{LN6}.
Thus, the structure of the CF wave functions is remarkably similar to the exact wave functions, the only difference being the incomplete center of mass factors.\footnote{For comparison, note that one may trivially rewrite the exact single vortex wave function as
$
\psi_{[N,N]}^{ex} = \sum_{i=1}^N   \left( z_i - Z \right) \cdot \prod_{j \neq i} \left( z_j - Z \right),
$
where the product part obviously approaches $ \psi_{[N-1,N-1]}^{(i)ex} =  \prod_{j \neq i} \left( z_j - Z ^{(i)}\right)$ in the large $N$ limit.}
While suggestive, this is not a proof that the CF wave functions approach the exact ones in the large $N$ limit, since the number of coordinates 'missing' in the center of mass piece increases with $N$. Unfortunately, full analytic overlap calculations on these expressions seem quite intractable. Instead, we now revisit the problem within projection method II, where a partial analytic proof of the converging overlaps is possible.

\section{Single vortex -- projection II}
\label{sec:svp2}

\noi
In this section we reexamine the issue using a different LLL projection method, referred to as method II (see below for details). This is the most commonly used projection method in the CF quantum Hall literature, being numerically more manageable (considerably fewer derivatives) than full projection. For quantum Hall states, it is known to make basically no difference numerically which of these projections is used\cite{jainbook}. In our case, this is not entirely obvious; the increase and convergence of the overlaps is a subtle effect, and exactly how the derivatives act in the projection, may well make a difference.
In order to compare the composite fermion ansatz to the exact wave function, we here rewrite each of them in the form of an $N \times N$ determinant divided by a single Jastrow factor. This has the advantage that we can, in fact, compare the two determinants entry by entry. We will prove analytically that for each pair of entries, the overlaps typically decrease with $N$ for small numbers of particles, {\it but then start increasing monotonically, and all converge to 1 in the thermodynamic limit}. While looking very suggestive, this however does not necessarily prove that the same would be true for the entire determinant, {\it i.e.} for the overlaps between the full wave functions. This point will be discussed in more detail below.

\subsection{A useful identity}

\noi
We start this section by deriving a mathematical identity that will turn out very useful in the following. Consider the Slater determinant
\be
 A^{(k)} [ z_1, ... z_N ]  \equiv
\left| \begin{array}{cccc}
1 & 1 &  ... & 1 \\
z_1 & z_2 & ... & z_N \\
... & ... & ... & ... \\
z_1^{k-1} & z_2^{k-1} & ... & z_N^{k-1} \\
z_1^{k+1} & z_2^{k+1} & ... & z_N^{k+1} \\
... & ... & ... & ... \\
z_1^{N}& z_2^{N} & ... & z_N^{N}
 \end{array} \right|,
\ee
which includes the Vandermonde determinant as the special case $k=N$. In general, being fully antisymmetric, $A^{(k)}$ is always divisible by a Vandermonde determinant,
\be
 A^{(k)} [ z_1, ... z_N ] =  {f}_{N-k}^S(\{ z_i \}) \cdot \prod_{i<j} (z_i - z_j),
\label{task2}
\ee
where ${f}_{N-k}^S$ is some fully symmetric polynomial of degree $N-k$. 
In fact, ${f}_{N-k}^S$ is, by definition, the Schur polynomial\cite{trudy} $s_{\lambda}(\{ z_i \})$ where the partition $\lambda = (1,1, ...1, 0, 0, ...,0)$ contains $(N-k)$ 1's followed by $k$ zeroes. From the Jacobi-Trudy identity\cite{trudy}, which prescribes how to express Schur polynomials in terms of fundamental symmetric polynomials, it furthermore follows that 
${f}_{N-k}^S(\{ z_i \})$ simply equals ${\cal S}_{N-k}(\{ z_i \})$, the fundamental symmetric polynomial of degree $N-k$ in the $N$ coordinates $\{z_i\}$.
Thus,
\be
{\cal S}_{N-k}(\{ z_i \}) = \frac{A^{(k)} [ z_1, ... z_N ] }{ \prod_{i<j} (z_i - z_j)}.
\label{TJ}
\ee
Some special cases of this identity have been discussed in the quantum Hall literature, e.g. \cite{macdonald, difrancesco}.

\subsection{Exact state}

\noi
Again, the exact ground state wave function at $L=N$ is given by ${\cal S}_N(\{ \tilde z_i \})$ where $\tilde z_i = z_i - Z$. Using Eq.\pref{TJ} with $k=0$, and $z_i$ replaced by $\tilde z_i$ (and noting that $(z_i - z_j) = (\tilde z_i - \tilde z_j) $), we can write this wave function in the form
\be
\psi_{L=N}^{ex} =  \frac{A^{(0)} [ \tilde z_1, ..., \tilde z_N]} { \prod_{i<j} (z_i - z_j)} \equiv   \frac{\det (M^{ex})} { \prod_{i<j} (z_i - z_j)}
\label{exdet}
\ee
with 
\be
M^{ex} &=&  \left( \begin{array}{cccc}
\zt_1^N & \zt_2^N & ... & \zt_N^N \\
... & ... & ... & ... \\
... & ... & ... & ... \\
\zt_1^{2} & \zt_2^{2} & ... & \zt_N^{2} \\
\zt_1& \zt_2 & ... & \zt_N
 \end{array} \right). 
 \label{Mex}
\ee
This form will turn out to be particularly convenient when comparing to the Jain wave function.

\subsection{Composite fermion ansatz}

\noi
The unprojected CF wave function for the single vortex was given in section \ref{sec:svp1}. To put this into a form comparable to Eq.\pref{exdet}, we use a standard trick due to Jain\cite{jainbook} to absorb {\it two} Jastrow factors into the determinant and perform the projection $\zb \rightarrow \p$ entry by entry,
\be
\psi_{(L=N)}^{CF}(\{ z_i \}) &=&  \left| \begin{array}{cccc}
z_1\prod_{k\neq 1}(z_1-z_k) & z_2 \prod_{k\neq 2}(z_2-z_k) & ... & z_N\prod_{k\neq N}(z_N-z_k) \\
1\cdot\prod_{k\neq 1}(z_1-z_k) & 1\cdot\prod_{k\neq 2}(z_2-z_k) &  ... &1\cdot \prod_{k\neq N}(z_N-z_k) \\
\p_1 \prod_{k\neq 1}(z_1-z_k) & \p_2  \prod_{k\neq 2}(z_2-z_k)& ... & \p_N \prod_{k\neq N}(z_N-z_k)\\
... & ... & ... & ... \\
... & ... & ... & ... \\
\p_1^{N-2}\prod_{k\neq 1}(z_1-z_k) & \p_2^{N-2} \prod_{k\neq 2}(z_2-z_k) & ... & \p_N^{N-2}\prod_{k\neq N}(z_N-z_k)
 \end{array} \right|
 \cdot \frac{1}{ \prod_{i<j} (z_i - z_j)} \\ \nonumber \\
  &\equiv& \frac{\det (M^{CF})} { \prod_{i<j} (z_i - z_j)}.
  \label{MCF}
\ee
This is what we refer to as projection method II.\footnote{Note that despite the division by a Jastrow factor there is no risk of singularities in the resulting wave function. The determinant in Eq.\pref{MCF} is by construction fully antisymmetric, \ie has zeroes for all $z_i = z_j$, and thus a Jastrow factor can always be factorized from it.}

\subsection{Overlaps}

\noi
To summarize the above manipulations, we have put both the exact and CF wave function into the form 
\be
\psi = \frac{\det(M)}{\prod_{i<j} (z_i - z_j)}
\ee
where $M$ is an $N\times N$ matrix. Recalling that our goal is to shed light on to what extent these two wave functions appear to approach each other with increasing $N$, we thus compare the two determinants, and do so entry by entry: We analytically compute pairwise overlaps 
$\langle M_{ij}^{ex} | M_{ij}^{CF} \rangle$
between corresponding {\it individual entries} of the two determinants, and show that {\it each one of these overlaps converges towards 1 in the limit $N \rightarrow \infty$}. In other words, the two matrices $M^{ex}$ and $M^{CF}$ approach each other entry by entry in the thermodynamic limit. Now, since in this limit the size of the matrices themselves goes to infinity, this does not strictly speaking prove that the overlap between the full {\it determinants} does approach unity as well. We have, so far, not been able to come up with a conclusive argument to settle this mathematically highly nontrivial question.

In order to compute the overlap between LLL polynomials, one uses the orthonormality relation
\be
\langle z^m | z^n \rangle = \int d^2 z \, \bar z^m z^n e^{-|z|^2/2} = 2\pi \delta_{mn} 2^m m!
\label{ON}
\ee
for all coordinates.
The actual overlap calculations involve a rather large amount of straightforward but tedious algebra and combinatorics. The main steps are outlined in appendix \ref{app:ol}; here we merely state and analyze the results.

As a concrete example, let us start with the first row and compute the overlap between $M_{1\alpha}^{ex}$ and $M_{1\alpha}^{CF}$ as defined through \pref{Mex} and \pref{MCF}, respectively. The result will only depend on the row index, {\it i.e.} be independent of the second index $\alpha$, so we compute the overlap
\be
O_{1} \equiv \frac{\langle M_{11}^{ex}  | M_{11}^{CF}  \rangle}{\sqrt{\langle M_{11}^{ex}  | M_{11}^{ex}  \rangle  \langle M_{11}^{CF}  | M_{11}^{CF}  \rangle}},
\ee 
where the denominator has to be included since the wave functions are not a priori normalized. The result is
\be
O_{1}(N) = \left[  \left( 1 - \frac 1 N  \right)^{N-2} \cdot \sum_{k=0}^{N-1} \left( 1 - \frac{k}{N}\right) \, \frac{1}{k!} \right]^{-1/2}.
\label{o1}
\ee
Now, using the fact that
$\lim_{n \rightarrow\infty}\left( 1 + 1/n \right)^n = e$, and likewise, $\sum_{k=0}^{\infty} (1/k!) = e$, one finds that
\be
\lim_{N \rightarrow \infty} O_1(N) = \lim_{N \rightarrow \infty} \left[ e^{-1} \left( e - \frac{e}{N} \right)  \right]^{-1/2} = 1.
\ee
So the overlap between any two corresponding entries in the first row approaches unity in the thermodynamic limit. Similarly, one finds for the second row,
\be
O_2(N) = \left[  \left( 1 - \frac 1 N  \right)^{N-1} \cdot \sum_{k=0}^{N-1} \frac{1}{k!} \right]^{-1/2},
\label{o2}
\ee
which is again seen to converge towards unity in the limit $N \rightarrow \infty$. In fact, as shown in the appendix, one can derive an analytic expression for the overlaps in {\it any} row $\alpha +2$, where $\alpha =  1, \cdots, (N-2)$,
\be
O_{\alpha+2}(N) =  \left[  \left( 1 - \frac 1 N  \right)^{N-\alpha-1} \cdot \sum_{k=0}^{N-\alpha -1} \frac{1}{k!} 
\left(1 - \frac{k}{N-1}\right) \, \left(1 - \frac{k}{N-2}\right) \cdots \left(1 - \frac{k}{N-\alpha}\right)  \right]^{-1/2} \, .
\label{OL}
\ee
 We already examined some of the lowest rows. Let us also have a look at a couple of examples for the highest row indices. First, note that for the last ($N$th) row, the overlap equals unity for {\it all} $N$. This can be shown by evaluating Eq.\pref{OL} for $\alpha = N-2$, or by noting through direct calculation that $\p_1^{N-2}\prod_{k\neq 1}(z_1-z_k) = \sum_{i=2}^N (z_1 - z_i) = N(z_1 - Z) \propto \zt_1$. For row number $N-1$, {\it i.e.} $\alpha=N-3$, Eq.\pref{OL} gives, after some algebra,
\be
O_{N-1}(N) = \left[   \left( 1 - \frac 1 N  \right)^2 \cdot \left(  1 + \frac{2}{N-1} + \frac{1}{(N-1)(N-2)}  \right)   \right]^{-1/2},
\ee
which is again seen to approach unity as $N \rightarrow \infty$. Similarly,
\be
O_{N-2}(N) = \left[   \left( 1 - \frac 1 N  \right)^3 \cdot \left(  1 + \frac{3}{N-1} + \frac{3}{(N-1)(N-2)}  + \frac{1}{(N-1)(N-2)(N-3)}  \right)   \right]^{-1/2},
\ee
etc. 
\begin{figure}
{\psfig{figure=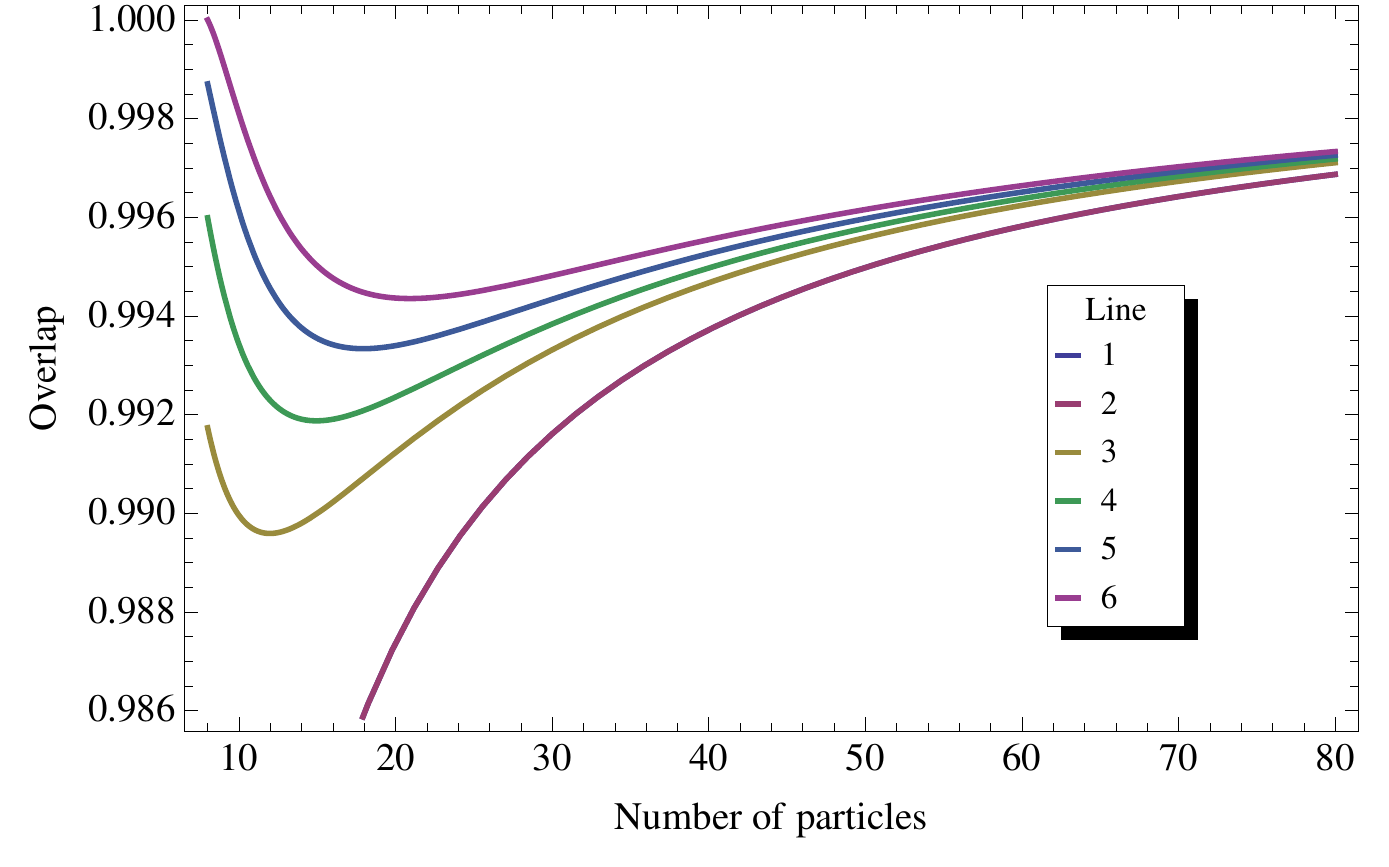, scale=0.6,angle=-0}}
\caption{Some row overlaps, given in Eqs.\pref{o1}-\pref{OL}, as functions of particle number. From below, rows 1 and 2 (too close to be resolved from each other), 3, 4, 5 and 6.}
\label{fig:ol}
\end{figure}
\begin{figure}
{\psfig{figure=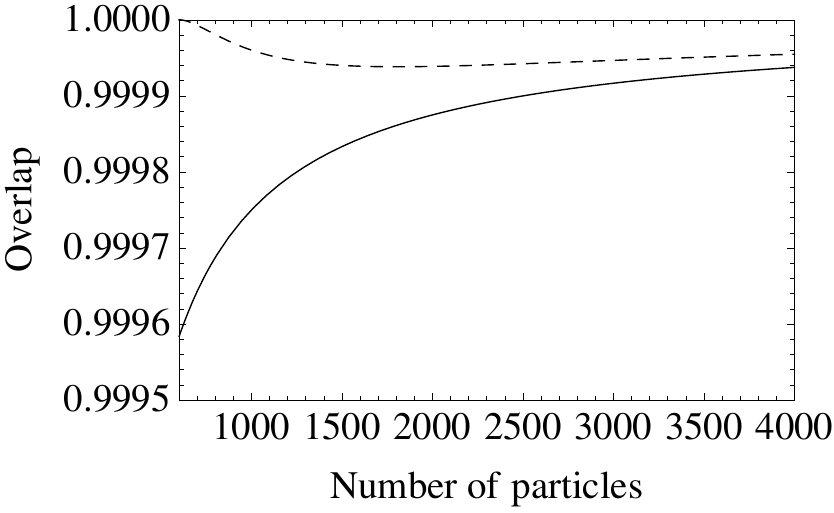, scale=0.9,angle=-0}}
\caption{Overlaps for first row (lower curve) and row number 600, for up to 4000 particles.}
\label{fig:ol2}
\end{figure}
%
  \begin{table}[htb]
  \begin{center}	  
    \begin{tabular}{| r | r | r | r | } \hline                  
     {\bf N} & {\bf L} & {\bf Projection I} & {\bf Projection II}       \\  \hline
     4 & 4  &  0.980 &  0.944       \\ \hline
     5 & 5  &  0.986 &  0.917       \\ \hline
     6 & 6  &  0.990 &  0.894       \\ \hline
     \end{tabular}
  \caption{Numerically computed overlaps between the full exact and CF wave functions at $L=N$ for the two projection methods discussed in the text. Note that the overlaps increase with $N$ for full projection (method I), but, for these small particle numbers, appear to decrease with $N$ for projection method II.}
  \label{table:LN}
  \end{center}
  \end{table}
Figures \ref{fig:ol} and \ref{fig:ol2} show some examples of overlaps versus particle number $N$. They illustrate that $O_{n+1} > O_{n}$ for given $N$, {\it i.e.} overlaps increase with increasing row index, and the first row, Eq.\pref{o1}, represents a lower bound of how fast the overlaps approach unity for growing $N$. As mentioned above, this does not provide a conclusive answer concerning the overlaps between the full wave functions. In fact, from table \ref{table:LN} we note that at {\it small} particle numbers, projection method II produces a {\it decreasing} overlap between the full wave functions as function of $N$, which may look discouraging at first sight. However, as figure \ref{fig:ol} illustrates, for system sizes this small, overlaps between the rows themselves actually tend to decrease rather rapidly as well, before passing a minimum and then increasing monotonically towards 1. It is thus fully conceivable that the overlaps between the full wave functions start increasing at larger $N$, too -- but the present analysis does not provide a conclusive answer to this question. It would be interesting to study this point numerically for rather large systems, say, 20-30 particles or more. Incidentally, if it did turn out that the full overlaps do not converge towards unity at large $N$, this would be a very rare example of a system where the choice of projection method makes an important qualitative difference to the behaviour of the CF wave functions.

\section{Lower angular momenta}
\label{sec:LlN}

\noi
The techniques developed for the single vortex in the previous sections, can be immediately transferred to the yrast states below the single vortex, $L < N$, and we shall again do this for the two projection methods separately. One apparent complication arising here is that for $L < N$, there will be more than one candidate CF Slater determinant for each state, the number of candidates increasing with decreasing $L$. So naively one would expect the ground state to be some linear superposition of these. However, remarkably, it turns out that after projection and antisymmetrization, the different CF candidates for a given $L$ and $N$ typically result in identical wave functions, leaving us with only one, representative, candidate state. 
%

\subsection{Projection I}

\noi
We start by discussing the case $L = N-1$, before generalizing to lower angular momenta. To construct this state using the CF scheme, one finds two candidate Slater determinants for the ground state\footnote{CF candidates for ground states have to be {\it compact}\cite{jainbook}. This means that the particles in any given CF Landau level occupy consecutive angular momentum states without any 'holes', starting from $l_n^{min} = -n$.}:
\be
\Phi_{[N-1,N]}(\{ z_i, \bar z_i \}) =  \left| \begin{array}{cccc}
z_1 & z_2 & ... & z_N \\
1 & 1 &  ... & 1 \\
\p_1 & \p_2 & ... & \p_N \\
... & ... & ... & ... \\
... & ... & ... & ... \\
\p_1^{N-3} & \p_2^{N-3} & ... & \p_N^{N-3} \\
\p_1^{N-1}& \p_2^{N-1} & ... & \p_N^{N-1}
 \end{array} \right|.
\label{LNm1}
\ee
and
\be
\tilde\Phi_{[N-1,N]}(\{ z_i, \bar z_i \}) =  \left| \begin{array}{cccc}
1 & 1 &  ... & 1 \\
z_1\p_1 & z_2\p_2 & ... & z_N\p_N \\
\p_1 & \p_2 & ... & \p_N \\
... & ... & ... & ... \\
... & ... & ... & ... \\
\p_1^{N-2} & \p_2^{N-2} & ... & \p_N^{N-2} 
 \end{array} \right|.
\ee
Manipulating these, one can show that these two candidates, when acting on a single Jastrow factor, produce identical wave functions.\footnote{We have checked this explicitly for up to six particles; since the corresponding manipulations will be completely analogous for higher $N$, we believe this to be true in general.} Thus, it is sufficient to consider one of them, and we shall choose the first, Eq.\pref{LNm1}. Proceeding similarly to the steps of Eqs. \pref{LN4a} - \pref{LNe}, we expand the Slater determinant \pref{LNm1} by the row with the highest number of derivatives to cast the CF wave function in the form
\be
\psi_{[N-1,N]}^{CF} = \sum_{i=1}^N \, \left| \begin{array}{cc}
1 &  ... \\
z_j & ... \\
\p_j & ... \\
... & ...  \\
... & ...  \\
\p_j^{N-3} & ... 
 \end{array} \right|^{(i)}
 \prod_{\alpha < \beta}^{~~~~(i)} (z_{\alpha} - z_{\beta}) \cdot \p_i^{N-1} \, \prod_{k \neq i}^{N-1} (z_i - z_k).
\ee
Noting that the $N-1$ derivatives simply turn the last part into a constant, and that the first part has the form of a sum over $(N-1)$-particle single vortex states (discussed in section \ref{sec:svp1}) with particle $i$ omitted, we thus have
\be
\psi_{[N-1,N]}^{CF} = \sum_{i=1}^N \, \psi_{[N-1,N-1]}^{(i)CF} .
\ee
So the CF yrast state at $L = N-1$ is simply a symmetrized sum over $L=N$ states with one particle less; these, in turn, are known explicitly from section \ref{sec:svp1}.

Similar considerations apply for lower $L$, where there will be even more CF candidates for each state. Although we have not bothered to perform a complete, systematic study for all possible cases, we have checked for a large class of these candidates that they again produce identical wave functions (or zero). Since the analytic structures of the wave functions at lower $L$ are very analogous to those encountered above, we believe this to be true in general\footnote{See also the discussion in Ref.\cite{korslund} on the lowest angular momentum states, $L=2$ and 3. For these cases it is known that the CF scheme produces the exact wave functions for all $N$, and that paper shows how all candidate CF Slater determinants ($N-1$ candidates for $L=2$, $N-2$ for $L=3$) reduce to identical wave functions.}. 
We thus, again, concentrate on one representative candidate, most conveniently the one whose Slater determinant has rows 
$\left[ z, 1, \p, ... , \p^{N-k-2}, \p^{N-k}, ..., \p^{N-1}\right]$, the generalization of \pref{LNm1}. 
The above calculation then immediately generalizes to give the result
\be
\psi_{[N-k,N]}^{CF} = \sum_{i=1}^N \, \psi_{[N-k,N-1]}^{(i)CF} .
\label{Nkp1}
\ee
Thus, using this expression iteratively, the CF yrast state at any $L = N-k$ can be expressed as a symmetrized (multiple) sum over single vortex states with $N-k$ particles, which are known from section \ref{sec:svp1}. For example,
\be
\psi_{[N-2,N]}^{CF} = \sum_{i=1}^N \, \psi_{[N-2,N-1]}^{(i)CF}  = \sum_{i=1}^N\sum_{j \neq i}^{N-1} \, \psi_{[N-2,N-2]}^{(ij)CF} ,
\ee
and analogously for smaller $L$. Note again the striking similarity to the structure of the corresponding exact yrast wave function \pref{TI1}.
Table \ref{table:LN12} shows some numerically computed overlaps between the exact and yrast wave functions for the $L=N, N-1,$ and $N-2$ states. We note that in all cases, the overlaps indeed increase with the number of particles, just as for the single vortex state\footnote{This further justifies our choice of one, representative CF state. Even if there were other, non-equivalent CF candidates, these would not contribute to the CF ground state in a limit where our representative state approaches the exact ground state.}. At the same time, overlaps tend to decrease as one moves from the single vortex towards lower angular momentum states at fixed $L$ or $N$. This is perhaps not very surprising, given that the state at $L=N-k$ can be expressed through single vortex states at a {\it smaller} number ($N-k$) of particles. The results of this section thus strongly suggest that the same kind of behaviour will be found for all yrast states in the interval $4 \leq L \leq N$ (not including $L=2,3$ here since there the CF wave function is known to be exact for all $N$\cite{viefers1}).

\bigskip

  \begin{table}[htb]
  \begin{center}	  
    \begin{tabular}{| r || r | r | r | } \hline                  
     {\bf L} & {\bf L=N} & {\bf L=N-1} & {\bf L=N-2}       \\  \hline
     4 &   0.980 &  0.953 & 0.926       \\ \hline
     5 &   0.986 &  0.967 & 0.945     \\ \hline
     6 &   0.990 &  0.973 & 0.954      \\ \hline
     \end{tabular}
  \caption{Numerically computed overlaps (projection I) between exact and CF wave functions at $L=N$, $N-1$, and $N-2$ for different values of $L$ (so the values of $N$ range from 4-6, 5-7, and 6-8, respectively in the three columns). Note that the overlaps increase with $N$ in all three cases. For fixed $L$, they tend to decrease as one moves away from the single vortex towards lower angular momenta.}
  \label{table:LN12}
  \end{center}
  \end{table}

\subsection{Projection II}

\noi
The analysis of the $L < N$ yrast states with projection II will be very analogous to that of section \ref{sec:svp2} for the single vortex. The aim of this section is to demonstrate how we can, again, express both the exact and CF states as a determinant divided by Jastrow factor, and compare the determinants entry by entry. 
Considering first the exact ground state at $L = N-k$, Eq.\pref{TI1}, we can again use Eq.\pref{TJ}, with general $k$, and $z_i$ replaced by $\tilde z_i$, to write
\be
\psi_{[N-k,N]}^{ex} = {\cal S}_{N-k}(\{ \zt_i \}) =  \frac{A^{(k)} [ \tilde z_1, ..., \tilde z_N]} { \prod_{i<j} (z_i - z_j)} \equiv   \frac{\det (M_{[N-k,N]}^{ex})} { \prod_{i<j} (z_i - z_j)}
\label{exdet2}
\ee
with 
\be
M_{[N-k,N]}^{ex} &=&  \left( \begin{array}{cccc}
\zt_1^N & \zt_2^N & ... & \zt_N^N \\
... & ... & ... & ... \\
\zt_1^{k+1} & \zt_2^{k+1} & ... & \zt_N^{k+1} \\
\zt_1^{k-1} & \zt_2^{k-1} & ... & \zt_N^{k-1} \\
... & ... & ... & ... \\
\zt_1& \zt_2 & ... & \zt_N \\
1 & 1 &  ... & 1 
 \end{array} \right). 
 \label{Mex2}
\ee
For the CF yrast state at $L = N-k$, we again choose to study the representative Slater determinant discussed in connection with Eq.\pref{Nkp1}. This immediately leads to the following analog of Eq.\pref{MCF},
\be
\psi_{[N-k,N]}^{CF} &=&  \left| \begin{array}{cccc}
z_1\prod_{k\neq 1}(z_1-z_k) & z_2 \prod_{k\neq 2}(z_2-z_k) & ... & z_N\prod_{k\neq N}(z_N-z_k) \\
1\cdot\prod_{k\neq 1}(z_1-z_k) & 1\cdot\prod_{k\neq 2}(z_2-z_k) &  ... &1\cdot \prod_{k\neq N}(z_N-z_k) \\
... & ... & ... & ... \\
\p_1^{N-k-2}\prod_{k\neq 1}(z_1-z_k) & \p_2^{N-k-2} \prod_{k\neq 2}(z_2-z_k) & ... & \p_N^{N-k-2}\prod_{k\neq N}(z_N-z_k) \\

\p_1^{N-k}\prod_{k\neq 1}(z_1-z_k) & \p_2^{N-k} \prod_{k\neq 2}(z_2-z_k) & ... & \p_N^{N-k}\prod_{k\neq N}(z_N-z_k) \\
... & ... & ... & ... \\
\p_1^{N-1}\prod_{k\neq 1}(z_1-z_k) & \p_2^{N-1} \prod_{k\neq 2}(z_2-z_k) & ... & \p_N^{N-1}\prod_{k\neq N}(z_N-z_k) 
 \end{array} \right|
 \cdot \frac{1}{ \prod_{i<j} (z_i - z_j)} \\ \nonumber \\
  &\equiv& \frac{\det (M_{[N-k,N]}^{CF})} { \prod_{i<j} (z_i - z_j)}.
  \label{MCF2}
\ee
Note that the last row of $M_{[N-k,N]}^{CF}$ reduces to a row of identical constants, thus being identical (up to an overall constant) to the last row of $M^{ex}_{[N-k,N]}$. All other rows have entries we already encountered in our discussion of the single vortex in section \ref{sec:svp2}. From the results of that section, we can thus immediately conclude that {\it between each pair of entries of these two determinants, for all $4 \leq L \leq N$, the overlaps converge to unity in the thermodynamic limit.} In other words, all yrast states below the single vortex behave in a way very analogous to that found for $L=N$.

\section{Conclusions}
\label{sec:concl}

\noi
The aim of this paper was to shed light on the surprising fact that trial CF wave functions for yrast states at and below the single vortex increase with system size, and appear to converge to the exact ground states in the thermodynamic limit, as suggested by numerics using projection method I. The ultimate goal would of course be a full derivation in terms of analytic expressions for the overlaps between CF and exact wave functions for arbitrary $N$. One might also wish to study analytic expressions for the eigenenergies of the CF states as functions of $L$ and $N$ and compare to their exact counterparts\cite{papenbrock01}. Both tasks turn out to be very challenging. Here, we reported progress in this direction by comparing analytically the form of the CF and exact yrast wave functions for the two most common projection methods, and pointing out their striking similarities. In particular, our analytical overlap calculations (in projection method II) for individual matrix entries should be viewed as suggestive evidence of the converging overlaps of the full wave functions, but do not provide a conclusive answer -- the two main 'problems' being that the sizes of the matrices at hand themselves go to infinity in the thermodynamic limit, and that the increase of the individual row overlaps is non-monotonic. It would thus be interesting to study the full overlaps for projection method II numerically for large systems. If they do {\it not} converge to unity, this would be one of very few known cases where the choice of projection method in the CF construction makes an important qualitative difference. One may also hope that some of the identities and manipulations pointed out in this paper, may turn out useful in other contexts, where exact analytic wave functions are not available. Finally, the calculations in this paper were done entirely in disk geometry; it would be of interest to reexamine the issues discussed here in, {\it e.g.}, spherical geometry.


\vskip 4mm
\noi {\bf Acknowledgement}: We thank Stellan \"Ostlund, Steve Simon, S\o ren Eilers, Hans Hansson, and Bernhard Mehlig for 
enlightening discussions. This work was supported by the Norwegian Research Council.


\appendix
\section{Analytic overlap calculations}
\label{app:ol}

\noi
In this appendix, we derive the expression for the overlap given by
Eq.\pref{OL} for a general number of particles $N$ and a given row
$\beta$. To prove this result we extensively use the binomial and
multinomial expressions :
\begin{eqnarray}
  \label{eq:a1}
  (a - b)^\alpha = \sum^\alpha_{i=0} (-1)^{\alpha - i}\left(\begin{array}{c} \alpha \\ i \end{array}\right) a^i b ^{\alpha-i}\\
  \left(\sum^n_i a_i\right)^\alpha = \sum_{|{\bf k}|=\alpha}  \left(\begin{array}{c} \alpha \\ {\bf k} \end{array}\right) \prod^n_i a^{k_i}_i,
\end{eqnarray}
with 
\begin{eqnarray}
  \label{eq:a2}
  \left(\begin{array}{c} \alpha \\ i \end{array}\right) = \frac{\alpha !} {(\alpha-i) ! i!}\\
  \left(\begin{array}{c} \alpha \\ {\bf k} \end{array}\right)=\frac{\alpha !}{\prod_i k_i !},
\end{eqnarray}
where ${\bf k}=(k_1,\cdots, k_n)$ and $|{\bf k}|=\sum^n_i k_i =
\alpha$. Since we are interested in the overlap between the matrix entries
$M^{\text{ex}}_{\beta i}$ and $M^{\text{CF}}_{\beta i}$, defined through Eqs. \pref{Mex} and \pref{MCF} respectively, we first
calculate the norm of each polynomial. The norm of
$M^{\text{ex}}_{\beta i}$ is given by (consistently suppressing factors of $2\pi$ from here on)
\begin{eqnarray}
  \label{eq:a3}
  \left<M^{\text{ex}}_{\beta i} |M^{\text{ex}}_{\beta i}\right> &=& \left<\left.\left(z_i-\frac1N \sum^N_{\alpha=1} z_\alpha\right)^\beta\right|\left({z}_i-\frac1N\sum^N_{\alpha=1} z_\alpha\right)^\beta\right>\label{eq:a3-1}\\
  &=&\left(\frac{N-1}N\right)^{2\beta}\sum^\beta_{\gamma=0} 2^\gamma\gamma !   \left(\begin{array}{c}\beta\\ \gamma\end{array}\right)2 \left(\frac{1}{N-1}\right)^{2(\beta-\gamma)}\sum_{\substack{|{\bf k}|=\beta-\gamma\\|{\bf k}^\prime|=\beta-\gamma}}\left(\begin{array}{c}\beta-\gamma\\ {\bf k}\end{array}\right)\left(\begin{array}{c}\beta-\gamma\\ {\bf k}\end{array}\right)\prod_{\alpha}\prod_{\omega} \left<z^{k_\alpha}_\alpha| z^{k^\prime_\omega}_\omega\right>\label{eq:a3-2}\\
  &=&2^\beta\left(\frac{N-1}N\right)^{2\beta}\sum^\beta_{\gamma=0}\gamma !   \left(\begin{array}{c}\beta\\ \gamma\end{array}\right)2 \left(\frac{1}{N-1}\right)^{2(\beta-\gamma)}(\beta-\gamma)!\sum_{|{\bf k}|=\beta-\gamma}\left(\begin{array}{c}\beta-\gamma\\ {\bf k}\end{array}\right)\label{eq:a3-3}\\
  &=&2^\beta \beta! \left(\frac{N-1}N\right)^{2\beta}\sum^\beta_{\gamma=0}\left(\begin{array}{c}\beta\\ \gamma\end{array}\right)\left(\frac{1}{N-1}\right)^{\beta-\gamma}\label{eq:a3-4}\\
  &=&2^\beta\beta!  \left(\frac{N-1}N\right)^\beta\label{eq:a3-5}.
\end{eqnarray}
In the first step, we applied the above binomial expansion and
the orthogonality relation \pref{ON} for the variables $z_i$. Then, Eq.~(\ref{eq:a3-3}) can be obtained
after expanding the remaining sum with the multinomial expression and
the orthogonality relation. Further simplifications give rise to
Eq.~(\ref{eq:a3-5}).

To calculate the norm of $M^{\text{CF}}_{\beta i}$, we first need to
evaluate the following expression:
\begin{eqnarray}
  \label{eq:a4}
  \partial^{\beta}_{z_1}\prod^{N}_{\beta\neq 1}\left(z_1-z_\beta\right)&=&\partial^{\beta}_{z_1}\sum^{N-1}_{\alpha=0} (-1)^{N-\alpha-1}\mathcal{S}_{N-\alpha-1}(\{z_i,i\neq 1\})z^\alpha_1 \\
  &=&\sum^{N-1}_{\alpha=\beta} (-1)^{N-\alpha-1}\mathcal{S}_{N-\alpha-1}(\{z_i,i\neq 1\}) \frac{\alpha!}{(\alpha-\beta)!} z_1^{\alpha-\beta}\\
&=&\sum^{N-\beta-1}_{\alpha=0}(-1)^{N-\beta-\alpha-1}\frac{(\alpha+\beta)!}{\alpha!}\mathcal{S}_{N-\beta-\alpha-1}(\{z_i,i\neq 1\})  z_1^\alpha,
\end{eqnarray}
where $\mathcal{S}_\alpha(\{z_i\})$ are the elementary symmetric
polynomials of degree $\alpha$. The norm of Eq.~(\ref{eq:a4}) can then easily
be deduced by using the orthogonality relation \pref{ON}. It
is given by
\begin{equation}
  \label{eq:a5}
  \left<\left.\partial^{\beta}_{z_1}\prod^{N}_{\beta\neq 1}\left(z_1-z_\beta\right)\right|\partial^{\beta}_{z_1}\prod^{N}_{\beta\neq 1}\left(z_1-z_\beta\right)\right> = 2^{N-\beta-1}(N-1)!\sum^{N-\beta-1}_{\alpha=0}\frac{(\alpha+\beta)!}{\alpha!(N-\beta-\alpha-1)!}.
\end{equation}
Finally the overlap is given by
\begin{eqnarray}
  \label{eq:a6}
  \left<\left. \partial^{\beta}_{z_1}\prod^{N}_{\lambda\neq 1}\left(z_1-z_\lambda\right) \right|\left(z_1-\frac{1}{N}\sum^N_{i=1}z_i\right)^{N-\beta-1}\right>&=&\left(\frac{N-1}{N}\right)^{N-\beta-1}\sum^{N-\beta-1}_{\substack{\lambda=0\\\alpha=0}}(-1)^{\alpha-\lambda} \frac{(\alpha+\beta)!}{\alpha!} \left(\begin{array}{c}N-\beta-1\\ N-\beta-\lambda-1\end{array}\right)\left<z_1^\alpha| z^\lambda_1\right>\nonumber \\
  &\times& \left(\frac{1}{N-1}\right)^{N-\beta-\lambda-1}\left<\mathcal{S}_{n-\beta-\alpha-1}(\{z_i,i\neq 1\})|\left(\sum^N_{i=2}z_i\right)^{N-\beta-\lambda-1}\right>\label{eq:a6-2}\\
  &=&\left(\frac{N-1}{N}\right)^{N-\beta-1}\sum^{N-\beta-1}_{\lambda=0}2^\lambda(\lambda+\beta)!\left(\begin{array}{c}N-\beta-1\\ N-\beta-\lambda-1\end{array}\right)\nonumber\\
  &\times&
  \left(\frac{1}{N-1}\right)^{N-\beta-\lambda-1}\left< \left.\mathcal{S}_{N-\beta-\lambda-1}(\{z_i,i\neq 1\})\right|\left(\sum^N_{i=2}z_i\right)^{N-\beta-\lambda-1}\right>\label{eq:a6-3}\\
  &=&2^{N-\beta-1}\left(\frac{N-1}{N}\right)^{N-\beta-1}\sum^{N-\beta-1}_{\lambda=0}(N-\beta-\lambda-1)!(\lambda+\beta)!\nonumber\\
  &\times&\left(\begin{array}{c}N-\beta-1\\ N-\beta-\lambda-1\end{array}\right)\left(\begin{array}{c}N-1\\ N-\beta-\lambda-1\end{array}\right) \left(\frac{1}{N-1}\right)^{n-\beta-\lambda-1}\\
  &=& 2^{N-\beta-1} (N-1)!\label{eq:a6-4}
\end{eqnarray}
To obtain Eq.~(\ref{eq:a6-3}), we first applied the orthogonality
relation to the variable $z_1$, then expanded the remaining multinomial
expression, only keeping terms involving $k_\alpha = 0$ or $1$
(as other terms do not give any contributions to the overlap). Combining
Eq.~(\ref{eq:a6-4}), (\ref{eq:a5}) and (\ref{eq:a3-5}), we thus find that the correctly normalized
overlap for any given row $\beta+2$ is given by
\begin{equation}
  \label{eq:7}
O_{\beta+2} \equiv \frac{ \left<M^{\text{ex}}_{\beta i} |M^{\text{CF}}_{\beta i}\right>}{\sqrt{\left<M^{\text{ex}}_{\beta i} |M^{\text{ex}}_{\beta i}\right>\left<M^{\text{CF}}_{\beta i} |M^{\text{CF}}_{\beta i}\right>}}
 = \frac{1}{\sqrt{\left(\frac{N-1}{N}\right)^{N-\beta-1}\sum^{N-\beta-1}_{\alpha=0}\frac{1}{(N-\beta-\alpha-1)!}\prod^\beta_{\lambda=1}\frac{\alpha+\lambda}{N-\lambda}}}.
\end{equation}
By reorganizing the factorials and products in the denominator of Eq.\pref{eq:7}, the sum over $\alpha$ can be recast to give the final expression \pref{OL}.

The overlaps $O_1$ \pref{o1} and $O_2$ \pref{o2} for the first and the second rows (where the CF determinant does not contain derivatives) are calculated
separately, but the calculations follow the same general strategy as those demonstrated here. 


\vspace{-0.5cm}
\bibliographystyle{unsrt}

\vspace{0.5cm}
\end{document}